\documentclass[journal=jctcce,manuscript=article,layout=onecolumn]{achemso}

\usepackage{amsmath}
\usepackage{graphicx}
\usepackage{booktabs}

\usepackage[colorlinks=true, allcolors=blue]{hyperref}

\title{AceFF: A State-of-the-Art Machine Learning Potential for Small Molecules}
\author{Stephen E. Farr}
\affiliation{Acellera Labs, C Dr Trueta 183, 08005, Barcelona, Spain}
\author{Stefan Doerr}
\affiliation{Acellera Labs, C Dr Trueta 183, 08005, Barcelona, Spain}
\author{Alireza Moradzadeh}
\affiliation{NVIDIA, 2788 San Tomas Expressway, Santa Clara, California 95051, USA}
\author{Antonio Mirarchi}
\affiliation{Computational Science Laboratory, Universitat Pompeu Fabra, Barcelona Biomedical Research Park (PRBB), C Dr. Aiguader 88, 08003, Barcelona, Spain}
\author{Francesc Sabanés Zariquiey}
\affiliation{Acellera Labs, C Dr Trueta 183, 08005, Barcelona, Spain}
\author{Saee Paliwal}
\affiliation{NVIDIA, 2788 San Tomas Expressway, Santa Clara, California 95051, USA}
\author{Gianni De Fabritiis}
\email{g.defabritiis@acellera.com}
\affiliation{Computational Science Laboratory, Universitat Pompeu Fabra, Barcelona Biomedical Research Park (PRBB), C Dr. Aiguader 88, 08003, Barcelona, Spain}
\altaffiliation{Acellera, 38350 Fremont Blvd 203 Fremont CA, 94536 USA}
\altaffiliation{Instituci\'o Catalana de Recerca i Estudis Avan\c{c}ats (ICREA), Passeig Lluis Companys 23, 08010 Barcelona, Spain}

\begin{document}
\maketitle

\begin{abstract}
We introduce AceFF, a pre-trained machine learning interatomic potential (MLIP) optimized for small molecule drug discovery. While MLIPs have emerged as efficient alternatives to Density Functional Theory (DFT), generalizability across diverse chemical spaces remains difficult. AceFF addresses this by refining the TensorNet2 architecture and training it on a comprehensive dataset of drug-like compounds. This approach yields a force field that balances high-throughput inference speed with DFT-level accuracy. We introduce two models sharing the same architecture: \mbox{AceFF-2}, trained on our own dataset of small molecules and supporting the essential medicinal chemistry elements (H, B, C, N, O, F, Si, P, S, Cl, Br, I); and \mbox{AceFF-2-OMOL}, trained on a subset of OMol25 containing the AceFF-2 elements plus the common biomolecular ion elements Na and K. The models are explicitly trained to handle charged states. Validation against rigorous benchmarks, including complex torsional energy scans, molecular dynamics trajectories, batched minimizations, and tests of force and energy accuracy, demonstrates that AceFF is state-of-the-art for organic molecules in the accuracy and speed regime important for drug discovery. The AceFF-2 and AceFF-2-OMol model weights and inference code are available at \url{https://huggingface.co/Acellera/AceFF-2.0} and \url{https://huggingface.co/Acellera/AceFF-2-OMOL}.
\end{abstract}

\section{Introduction}

Atomistic simulations are fundamental tools for understanding materials and molecular systems at the atomic level. The accuracy and utility of these simulations critically depend on the choice of force field. Classical molecular mechanics (MM) force fields such as GAFF\cite{wang2004development,wang2006automatic}, CGenFF\cite{vanommeslaeghe2010charmm,vanommeslaeghe2012automation}, OpenFF\cite{qiu2021development,boothroyd2023development}, and AMBER\cite{maier2015ff14sb} offer high computational efficiency, but often fall short in predictive accuracy, particularly for diverse, drug-like molecules containing rare functional groups or where quantum effects and polarization play significant roles.  First-principles quantum mechanical (QM) methods, such as density functional theory (DFT), provide substantially higher accuracy but are prohibitively expensive for routine biomolecular simulations. 

Machine learning interatomic potentials (MLIPs) offer a promising alternative, achieving near-QM accuracy at a fraction of the computational cost:  an order of magnitude slower than MM methods, while being many orders of magnitude faster than QM approaches~\cite{Behler2007NNP,Bartok2010GAP,Smith2017ANI1,Unke2021MLFFReview}.
The current state-of-the-art MLIP architectures in terms of accuracy include equivariant message-passing graph neural network models such as MACE\cite{batatia_mace_2023}, Nequip\cite{batzner_e3-equivariant_2022}, AIMNet2 \cite{anstine2024aimnet2}, and TensorNet\cite{TensorNet}, to name a few\cite{duval2023hitchhiker}.  One of the most used models to date is ANI-2x\cite{ANI2x}, which employs a less accurate invariant architecture but is one of the most computationally efficient MLIPs available. A newer and more accurate group of pre-trained models are the MACE-OFF23 models with large, medium, and small variants available. These are significantly more accurate than ANI-2x, with the downside of being substantially slower~\cite{kovacs2023mace}. Other pre-trained models include: AIMNet2\cite{anstine2024aimnet2} which supports charged molecules, AceFF-1.0\cite{aceff_huggingface} based on the TensorNet architecture\cite{simeon_broadening_2025} supporting charge and spins, the Egret models, which are based on the MACE architecture\cite{mann_egret-1_2025}; the recent OrbMol\cite{orbmol_huggingface, rhodes2025orbv3atomisticsimulationscale, neumann_orb_2024}; and UMA-OMol25 models\cite{wood2026umafamilyuniversalmodels}. OrbMol and UMA are both trained on the recent OMol25 dataset~\cite{levine2025openmolecules2025omol25} and cover most of the periodic table, while being accurate in standard benchmarks, they are slow relative to more specialized models. A recent fast model is SO3LR~\cite{kabylda_molecular_2025} which incorporates semi-local interactions from an equivariant neural network with universal pairwise force fields designed for short-range repulsion,
long-range electrostatics, and dispersion interactions. 

It is currently not feasible to perform production MD using a full MLIP description of the entire molecular system. Instead, hybrid MLIP/MM schemes are necessary, where only the ligand is modeled with the MLIP and the remainder of the system uses a classical MM force field. It has been found that even with a simple mechanical embedding approach, MLIP/MM can provide increased accuracy for RBFE calculations\cite{sabanes2024enhancing, sabanes_zariquiey_quantumbind-rbfe_2025}. In our previous work, we were limited by the fact that ANI-2x supports only seven elements and only neutral molecules. This led us to create our own QM dataset and develop our first MLIP, AceFF-1.0\cite{aceff_huggingface}, which we applied in a large-scale RBFE benchmark, finding improved accuracy compared to traditional force fields and previous ANI-2x results\cite{sabanes_zariquiey_quantumbind-rbfe_2025}. AceFF-1.1 included a larger dataset and was released on HuggingFace\cite{aceff1.1_huggingface}. Both of these models are freely available under the Apache 2.0 license for academic and commercial use. In this paper, we report the latest updates to our model family, AceFF-2 and AceFF-2-OMOL which include both more training data and an improved model architecture, TensorNet2, and a faster implementation with optimized Nvidia warp kernels.  Additionally, we describe the various benchmarks we use to quickly assess model accuracy, making comparisons in accuracy with the existing MLIPs ANI-2x~\cite{ANI2x}, AIMNet2~\cite{anstine2024aimnet2}, OrbMol~\cite{orbmol_huggingface, rhodes2025orbv3atomisticsimulationscale}, UMA~\cite{levine2025openmolecules2025omol25}; semi-empirical methods g-XTB~\cite{froitzheim_g-xtb_2025}, GFN2-XTB~\cite{bannwarth_gfn2-xtbaccurate_2019}; and the MM forcefield GAFF2~\cite{wang2004development,wang2006automatic}. In summary, AceFF-2-OMOL is one of the fastest models with high accuracy, because of this we believe that it has immediate applications in drug discovery.

\section{Results}

\subsection{TensorNet2 architecture}

\begin{figure}
    \centering
    \includegraphics[width=\linewidth]{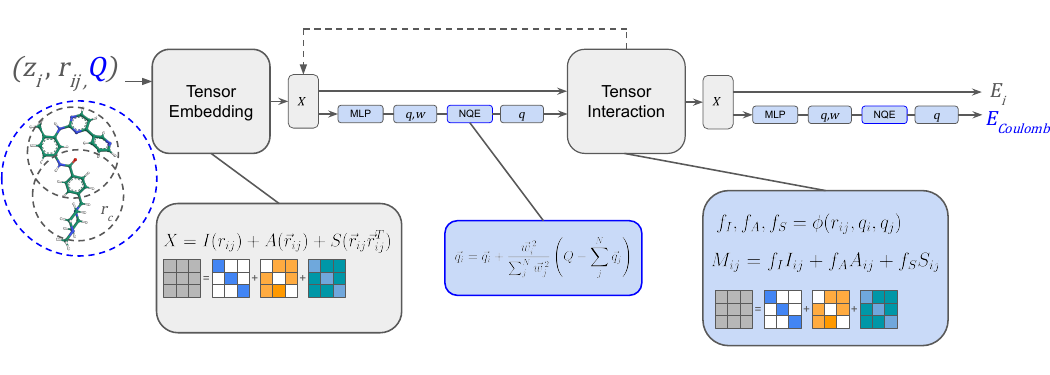}
    \caption{ The AceFF-2 architecture TensorNet2 builds upon TensorNet\cite{TensorNet} with the new components colored in blue. The main change is that after the tensor embedding and each tensor interaction block, a set of partial charges $q$ and weights $w$ are computed and undergo a neutral charge equilibration (NQE) procedure as done in AIMNet2\cite{anstine2024aimnet2}. The partial charges are then combined with edge features during the tensor interaction. The predicted charges are used to compute a Coulomb energy $E_\text{Coulomb}$ alongside the standard short-range node-wise MLIP energy term $E_i$. The dashed arrow represents where iteration occurs for more than one interaction layer. The pictured diagram corresponds to TensorNet2 1-layer. The grey dashed circles around the example molecule illustrate the short cutoffs used in the tensor embedding and interaction. The blue dashed circle represents the long-range of the total charge (a global feature), the NQE block, and the Coulomb interaction.}
    \label{fig:model}
\end{figure}

AceFF-1.x models were based on TensorNet\cite{TensorNet} models with charge and spin encodings\cite{simeon_broadening_2025}. In this work, we introduce a modified version of TensorNet with improved charge handling. The approach we take closely follows that of AIMNet2\cite{anstine2024aimnet2}: we include additional scalar features to mimic partial charges, perform neutral charge equilibration on them, and use them in a long-range Coulomb energy term. We name this version TensorNet2. An overview of the model architecture is shown in figure~\ref{fig:model}.

From an input of atomic numbers and coordinates $Z_i$,$R_i$, TensorNet learns a set of $3\times3$ matrices (rank-2 tensors) for each atom $X^{(i)}$ which are equivariant to rotation and translation. At each message-passing step, the $X^{(i)}$ are updated based on the neighboring atoms, making use of the fact that $X^{(i)}$
can be decomposed into scalar, vector, and
symmetric traceless components, $I$, $A$, $S$, respectively. After several message-passing updates, the squared Frobenius norms of the representations, which are invariant under O(3), are further processed by the neural network to predict energies, obtaining atomic forces via automatic differentiation. Moreover, a distinctive feature of the TensorNet message-passing architecture, specifically when interaction layers are active, is its natural transition from O(3) to SO(3) equivariance. This reduction in symmetry follows directly from how tensor–tensor interaction products are computed in the interaction layer, which mix even and odd parities and therefore break reflection equivariance while preserving full rotational equivariance. This behavior is especially valuable in drug discovery, where chirality is central: enantiomers are related by reflections but often differ dramatically in binding affinity and biological activity\cite{h2011significance}.

In the previous version of TensorNet\cite{simeon_broadening_2025}, total molecular charge $Q$ was included in a simple manner with no extra learned parameters. While this method effectively resolved the degeneracy problem for molecules sharing the same coordinates and atomic numbers but differing in $Q$, we observed limitations during extrapolation. Specifically, when tested on a set of larger charged molecules, the performance was significantly surpassed by the AIMNet2 architecture~\cite{anstine2024aimnet2}, which explicitly models and learns partial charges and Coulomb interactions.

We incorporate the AIMNet2-like neutral charge equilibration into TensorNet as follows, which is the main contribution to the TensorNet2 new architecture. At each message-passing layer, we introduce a charge prediction neural network that accepts the node features ($I^i, A^i, S^i$) as input and outputs a set of charge predictions ($\vec{q}_i$) and a corresponding set of weights ($\vec{w}_i$). It is important to note the inclusion of a feature dimension (a hyperparameter) to increase expressivity. These outputs are not required to represent physical charges but rather scalar features that we equate to partial charge hypotheses (similar to the approach in \cite{ple2025foundation}). A neutral charge equilibration step is then performed independently across each feature channel to ensure that the partial charges sum to the total molecular charge ($Q$).

\begin{equation}
\vec{q}_i \leftarrow \vec{q}_i + \frac{\vec{w}_i}{\sum^N_j \vec{w}_j} \left( Q - \sum^N_j \vec{q}_j \right),
\end{equation}
which is the same as in \cite{anstine2024aimnet2}.
The $q_i$ from each node are then used in the next message-passing update by extending the scalar edge features such that Equation 11 in \cite{TensorNet} becomes:
\begin{equation}
f^i_I, f^i_A, f^i_S = \phi(r_{ij})\,\mathrm{SiLU}\Big(\mathrm{MLP}\big(e^{\mathrm{RBF}}(r_{ij}) \oplus q_i \oplus q_j\big)\Big),
\end{equation}
where $\oplus$ denotes tensor concatenation.
The predicted $\vec{q}_i$ from all layers are subsequently used to compute Coulomb interactions, which is performed independently across each feature channel. The weighted mean of the resulting Coulomb energy is then calculated, where the weights are assigned such that the output from the final layer has the most significant impact on the total energy. Utilizing the charge predictions from the earlier layers in the final output calculation effectively serves as a regularization technique.

This procedure is an iterative refinement of the partial charges, in which the charges from each layer condition the features of the next, and is fundamentally different in style from alternative charge inclusion techniques such as the latent Ewald summation method~\cite{cheng_latent_2025}, which calculates an Ewald summation or a Coulomb-like interaction using node features derived from just the final node features. Another is the 4G-HDNNP approach~\cite{ko_fourth-generation_2021}, which employs neural networks to predict electronegativity and hardness. This prediction is followed by a full charge equilibration scheme, which involves solving a set of linear equations to find the set of partial charges that minimizes the electrostatic energy subject to the total charge constraint, which scales poorly with system size. The TensorNet2 neutral charge equilibration approach adds only a small overhead to training and inference speed. For example, a 1-layer TensorNet model with an embedding dimension of 128 and 32 radial basis functions has 535,681 parameters, while a 1-layer TensorNet2 model with the same embedding dimension, radial basis functions, and additionally a charge dimension of 16 has 685,413 parameters. The inference speeds of each model for a 1500 atom system (500 water molecules) are 100 steps/second and 75 steps/second, respectively. A parameter increase of 28\% and a speed decrease of 25\%.

\subsection{Software Optimizations}
\begin{figure}
    \centering
    \includegraphics[width=0.5\linewidth]{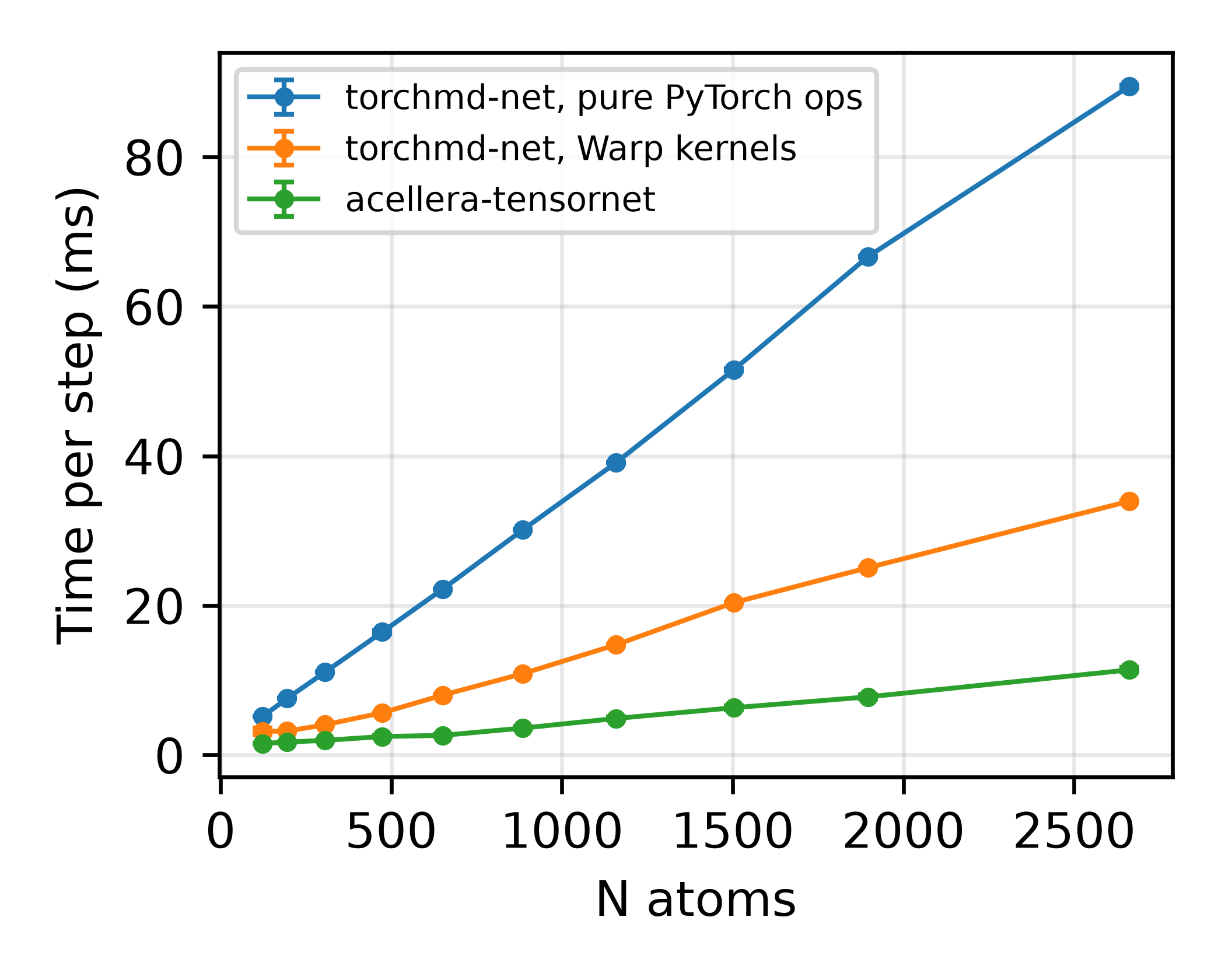}
    \caption{Molecular dynamics speed (time per step) of AceFF-2 for increasing number of atoms, using periodic water boxes with a reaction-field Coulomb cutoff of 1.2~nm (or half the box length for boxes smaller than 2.4~nm), on an RTX 4090. The blue line is torchmd-net with the pure PyTorch TensorNet ops, the orange line is torchmd-net version 3 with the Warp kernels, and the green line is the acellera-tensornet version.}
    \label{fig:old_v_new}
\end{figure}

With this model update, we have introduced software optimizations to the TensorNet models within the TorchMD-Net\cite{pelaez2024torchmd} package. These optimizations utilize custom NVIDIA warp kernels, implemented from the MatGL package\cite{Ko2025, matgl_repo}, for the primary bottleneck operations: the decomposition and composition between the $X$ and $I, A, S$ tensors, as well as the message-passing routines. This implementation results in a 3x speed-up for both inference and training. Figure \ref{fig:old_v_new} illustrates the molecular dynamics inference speed (time per step) as a function of system size, comparing the legacy code to the optimized version.
The updated code also achieves a 3x reduction in GPU memory usage. These improvements stem from a more efficient representation of the $I, A, S$ tensors. The original implementation stored these as three separate $3 \times 3$ tensors to facilitate PyTorch vectorization. Mathematically, however, $I, A,$ and $S$ represent a scalar, a vector, and a symmetric traceless tensor, requiring only 1, 3, and 5 values, respectively. The optimized kernels maintain these components in compact layouts of $(1,h)$, $(3,h)$, and $(5,h)$, where $h$ is the hidden dimension. This also avoids repeatedly converting between the $(h,3,3)$ and $(3,3,h)$ layouts required by the previous implementation. Message passing is performed by a fused kernel that processes the full molecular graph using two complementary compressed representations of the edge indices. One representation groups edges by destination node for efficient message aggregation during the forward pass, while the transposed representation groups them by source node for efficient gradient accumulation during the backward pass. These fused forward and backward traversals avoid separate gather, transformation, and scatter operations and eliminate the need to materialize large edge-sized intermediate tensors, substantially reducing memory traffic and Autograd overhead.

We also have a second implementation of TensorNet2, written entirely in C++ and CUDA exclusively for MD simulations (does not support training), which we call acellera-tensornet. It evaluates the whole model on the GPU without PyTorch, so the complete molecular dynamics loop can run with no per-step Python overhead. It is faster again than the Warp-accelerated TorchMD-Net implementation: on periodic water boxes on an RTX 4090 it is about three times faster at 2661 atoms (88.6 versus 29.5 steps/second) and about eight times faster than the pure PyTorch code (Figure~\ref{fig:old_v_new}).

\subsection{AceFF trained models}
We trained a 2-layer version of TensorNet2 on an internal QM dataset; we name this model AceFF-2. We also trained a model with the same architecture on a subset of OMol25\cite{levine2025openmolecules2025omol25}; we name this model AceFF-2-OMOL. The only architectural difference is the inclusion of an empirical ZBL repulsion term for the AceFF-2-OMOL model, as is typically done in other models~\cite{rhodes2025orbv3atomisticsimulationscale}. We used a charge embedding dimension of 16, a hidden dimension of 128, and 32 radial basis functions. The model was trained using the torchmd-train scripts that can be found in TorchMD-Net~\cite{pelaez2024torchmd}.  The model was trained on energy and force labels, without using any partial charge or dipole moment data. Optionally, one of the TensorNet2 charge channels could be trained on a QM-level-specific type of partial charge, e.g., MBIS, to facilitate an approach such as~\cite{semelak_advancing_2025}. Indeed, we expect the different charge channels could be trained simultaneously on multiple QM-derived partial charges—for example, the Omol25 dataset contains Mulliken and Lowdin charges~\cite{levine2025openmolecules2025omol25}.

\subsection{Accuracy on torsion scans}

\begin{figure}[t]
    \centering
    \includegraphics[width=\linewidth]{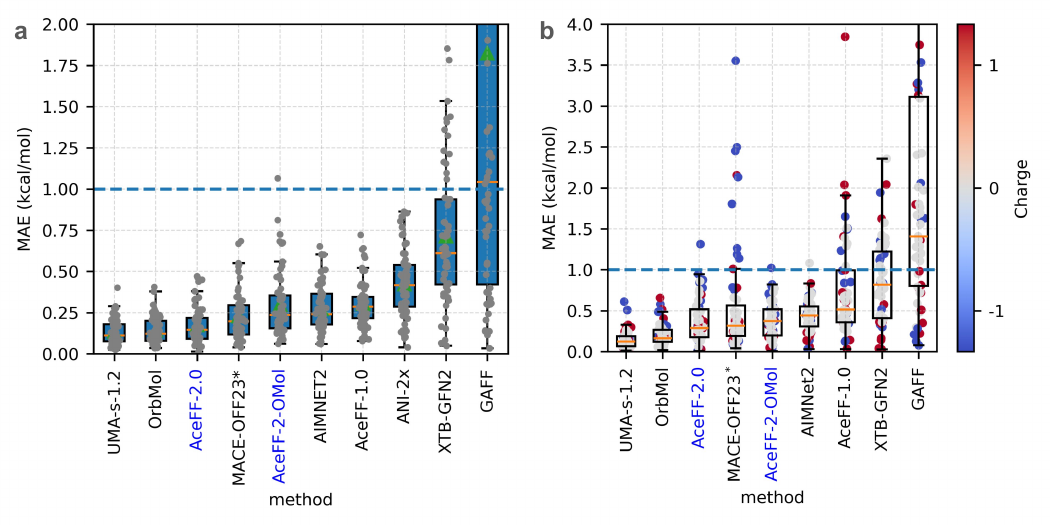}
    \caption{Torsion scans. {\bf a.} Sellers et al. torsion scan benchmark\cite{sellers2017comparison}. The orange lines are the median value. The methods are ordered from left to right by median MAE compared to the coupled cluster reference data. The y-axis has been intentionally capped at 2.0 kcal/mol to focus on the differences between the MLIP models. An image of one of the molecules with the torsion angle indicated is pictured. Values marked by * were taken from a previous publication~\cite{sabanes_zariquiey_quantumbind-rbfe_2025}. All other data points were recalculated. {\bf b.} Behara et al torsion scan benchmark~\cite{behara2024}. The orange lines are the median value. The methods are ordered from left to right by median MAE compared to the coupled cluster reference data. The scatter points are color-coded by the molecular charge. The y-axis has been intentionally capped at 4.0kcal/mol to focus on the differences between the MLIPs. Data points of methods marked with * were taken from the original data\cite{behara2024}, all other data points were recalculated. On both plots, a guiding line has been drawn at 1 kcal/mol. }
    \label{fig:torsion_scan}
\end{figure}

The first benchmark test we performed was Sellers et al torsion scan~\cite{sellers2017comparison}. This benchmark dataset contains torsion scans of 62 molecules done with different levels of QM. We take the most accurate method from the set, CCSD(T)/CBS, and use this as the reference to compare against. For all molecules in the test, we performed relaxed torsion scans (as done in\cite{smith_approaching_2019, smith_transforming_2018}). For comparison we did the same procedure with AceFF-2-OMOL, our previous AceFF-1.0, ANI-2x\cite{ANI2x} from TorchANI\cite{gao2020torchani}, AIMNet2\cite{anstine2024aimnet2}, UMA-s-1.2\cite{levine2025openmolecules2025omol25}, and the recent OrbMol-conservative\cite{orbmol_huggingface,rhodes2025orbv3atomisticsimulationscale,neumann_orb_2024}. We also did the scan with the semi-empirical method GFN2-xTB\cite{bannwarth_gfn2-xtbaccurate_2019} and the molecular mechanics force field GAFF\cite{wang2004development,wang2006automatic}. We excluded g-xTB because it does not have analytical gradients available yet. The MAE of each torsion scan relative to the reference scan is plotted in Figure~\ref{fig:torsion_scan}a as box plots with all data points shown.
We see that UMA-s-1.2 and OrbMol are the most accurate models, with AceFF-2 a close third. From left to right, MACE-OFF23, AceFF-2-OMol, AIMNet2, and AceFF-1.0 perform similarly, until ANI-2x, which is significantly worse. GFN2-xTB performs poorly compared to the MLIPs, and as expected, GAFF has very large errors.

We also performed a second torsion scan benchmark from Behara et al.~\cite{behara2024}. This torsion scan contains charged molecules, specifically [0,-1,+1]. The reference torsion drives were done with  MP2/heavy-aug-cc-pVTZ, and the reference single-point energies were labeled with CCSD(T)/CBS.  The benchmark differs from the Sellers et al torsion scan in that we do not perform any geometry optimizations. This benchmark tests only the energy error of the MLIP and not the force calculations. We did the single-point calculations for UMA-s-1.2, OrbMol, AceFF-2, AceFF-2-OMol, AceFF-1.0, AIMNet2, GFN2-xTB, and GAFF.

The MAE of each method is plotted in figure~\ref{fig:torsion_scan}b. The trend between models is approximately the same as the Sellers et al torsion scan. As expected, the biggest outliers in AceFF-1.0 are the charged molecules, and similarly for MACE-OFF23, which was designed for only neutral molecules. While UMA, OrbMol, and AceFF2 demonstrate a more balanced performance across both neutral and charged molecules, charged species still represent the most significant deviations for all three. Notably, AIMNet2 is the only model where the largest outlier is a neutral molecule.

\subsection{Accuracy on strained conformers: Wiggle150}

The Wiggle150 benchmark~\cite{brew_wiggle150_2025} consists of 150 highly strained conformations of adenosine, benzylpenicillin, and efavirenz, with reference energies calculated using DLPNO-CCSD(T)/CBS level of theory. To evaluate performance, we used the MLIP models to predict the relative energies (relative to the minimized geometry) for all 150 conformations. The resulting mean absolute error (MAE) and root mean squared error (RMSE) were then computed compared to the reference values.

Table~\ref{tab:wiggle} summarizes the Wiggle150 results for GAFF, GFN2-xTB, ANI-2x, g-xTB, AceFF-1.0, AceFF-2, AceFF-2-OMol, AIMNet2, OrbMol, and UMA-s-1.2. Data for ANI-2x, AIMNet2, and GFN2-xTB were taken from the original Wiggle150 publication~\cite{brew_wiggle150_2025}, while the remaining values were computed in-house using the respective publicly available models. OrbMol and UMA exhibit the highest accuracy. Notably, all MLIPs except ANI-2x outperform the semi-empirical g-xTB method, despite g-xTB being fitted to reproduce a similar level of theory ($\omega$B97M-V). Furthermore, AceFF-2 shows a significant improvement over AceFF-1.0, and AceFF-2-OMol improves on AceFF-2 again.

\begin{table}[t]
    \centering
    \begin{tabular}{@{}lcc@{}}
        \toprule
        \textbf{Method} & \textbf{MAE (kcal/mol)} & \textbf{RMSE (kcal/mol)} \\ 
        \midrule
        GAFF & 22.87 & 30.86 \\
        GFN2-XTB & 14.6 & 15.2 \\
        ANI-2X   & 4.41 & 5.41 \\ 
        g-XTB    & 3.85 & 4.76 \\
        AceFF-1.0 & 2.73 & 3.32 \\ 
        AIMNet2  & 2.39 & 3.13 \\ 
        AceFF-2 & 1.76  &  2.34 \\ 
        AceFF-2-OMol & 1.25 & 1.64 \\
        UMA-s-1.2 (omol task) & 0.92 & 1.23 \\
        OrbMol & 0.89 & 1.22 \\
        \bottomrule
    \end{tabular}
    \caption{Performance of MLIPs, two semi-empirical methods, and one MM force field on the Wiggle150 benchmark~\cite{brew_wiggle150_2025}.}
    \label{tab:wiggle}
\end{table}

\subsection{Drug-like molecules: Schrödinger ligands}

We created a test set that evaluates how well the model generalizes to larger, previously unseen molecules. These ligands are all larger than 30 atoms, which is the maximum size included in the AceFF DFT training dataset.  To create the Schrödinger ligand test set, we took the ligand structures from~\cite{ross2023maximal} available at \url{https://github.com/schrodinger/public_binding_free_energy_benchmark} and labeled them with forces and energies computed at the AceFF level of theory ($\omega$B97M-V/def2-TZVPPD). All ligands from the Merck, JACS, and charge-annihilation sets were included.

We used the conformations directly from the provided SDF files and performed DFT calculations with the same settings used to create the AceFF DFT dataset. Any conformations that failed to converge were discarded, resulting in a final set of 650 conformers. We predicted the force and energy using AceFF-1.0, AceFF-2, AceFF-2-OMol, AIMNet2, OrbMol, and UMA-s-1.2. For each method, we computed the force MAE per conformer and the energy absolute error. The energy is the total DFT energy; as such, we expect the AceFF models, which are trained on the same level of theory DFT, to have energy errors close to zero. The AIMNet2, OrbMol, and UMA models were trained on data generated by slightly different functionals, basis sets, and software. However, the force errors should be comparable between the models, as this is an intrinsic physical property. Therefore, we only show energy errors for the AceFF models.

\begin{figure}
    \centering
    \includegraphics[width=\linewidth]{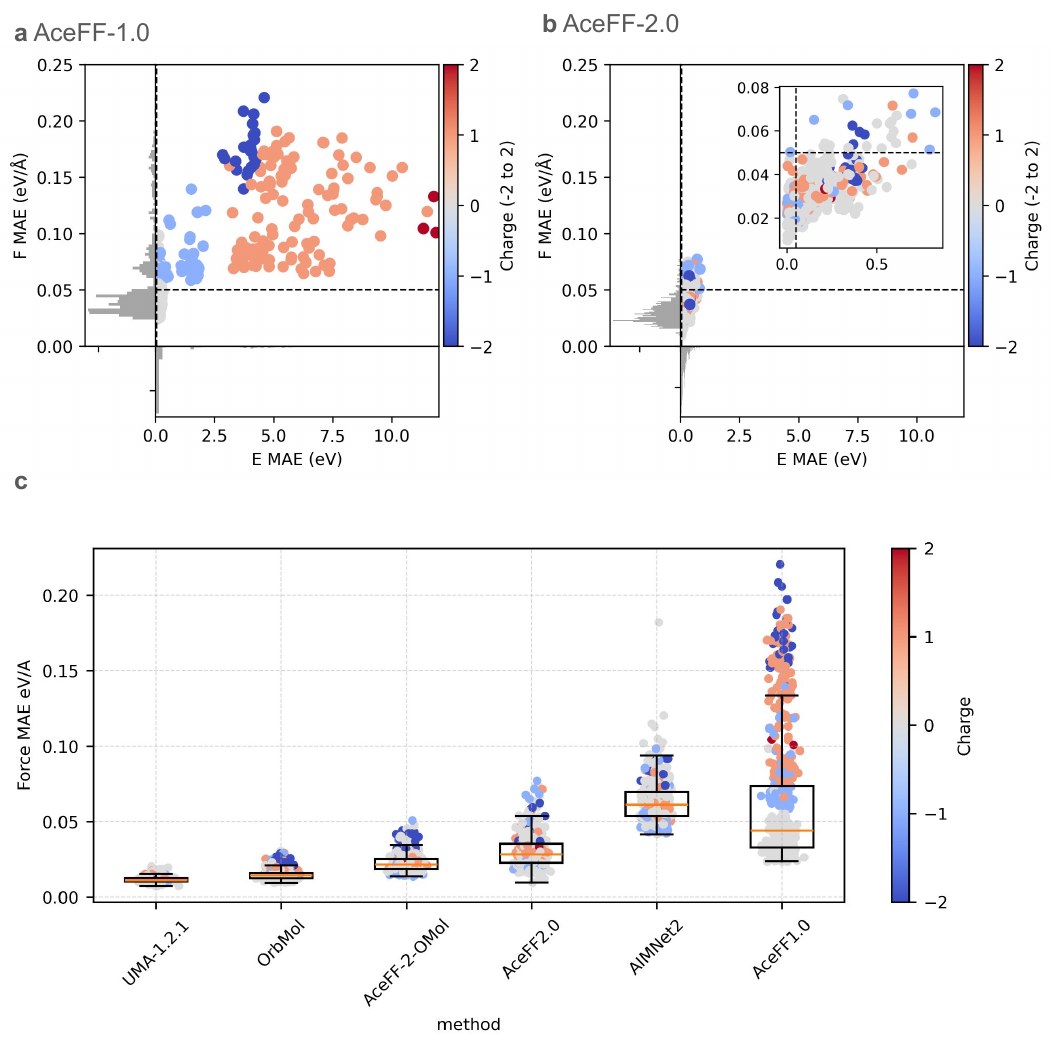}
    \caption{Schrödinger ligand test set evaluation. {\bf a.} Energy error vs force error for AceFF-1.0, color-coded by ligand total charge. {\bf b.} Energy error vs force error for AceFF-2, color-coded by ligand total charge. The axis range is the same in sub-plots a and b to aid comparison. The inset graph in sub-plot b shows a zoomed-in view of the data points. {\bf c.} Force errors for all tested models. The orange line is the median value. The data points are color-coded by charge.}
    \label{fig:schrodinger}
\end{figure}

The results are plotted in Figure~\ref{fig:schrodinger}a and~\ref{fig:schrodinger}b for AceFF-1.0 and AceFF-2, respectively, showing the force and energy errors. We have plotted guide lines at 0.05~eV/\AA{} and 0.05~eV, which indicate the approximate validation L1 force and energy loss that the models converge to. We see that, as expected, AceFF-1.0 performs very badly on charged molecules. AceFF-2 performs much better with the majority of force errors below 0.05~eV/\AA. The energy errors are not below 0.05~eV; however, these molecules are mostly larger (27 to 74 atoms) than any it was trained on.

The force errors for all models are plotted in Figure~\ref{fig:schrodinger}c. AIMNet2 has consistent performance between charged and neutral molecules, but is systematically above 0.05~eV/\AA. UMA-s-1.2 performs exceptionally well with force errors all below 0.02~eV/\AA, closely followed by OrbMol, which has all errors below 0.03~eV/\AA. AceFF-2-OMol sits between OrbMol and AceFF-2, with a median force error of about 0.02~eV/\AA{} and only a handful of charged outliers above 0.05~eV/\AA. We note that AceFF-2-OMol, OrbMol, and UMA-s-1.2 were all trained on the OMol25 dataset, which does contain molecules of this size (and may well contain these molecules or very similar ones in the training split). It is interesting to see that the most extreme outliers for AIMNet2 are neutral molecules.

\subsection{SPICE Test Set}

\begin{figure}
    \centering
    \includegraphics[width=\linewidth]{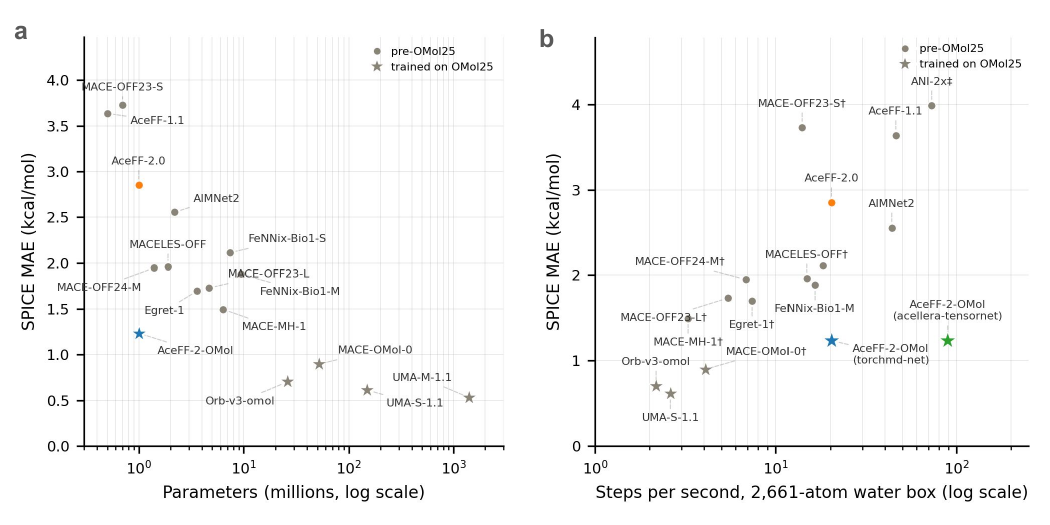}
    \caption{SPICE test set performance. \textbf{a.} Overall MAE plotted against the number of model parameters as in~\cite{eastman2026}. All gray data points are taken from~\cite{eastman2026}. Points are categorized into models trained on OMol25 (stars) and models trained on pre-OMol25 datasets (circles). \textbf{b.} Overall MAE plotted against simulation speed (steps per second for a 2,661-atom water box) as done in~\cite{eastman2026}, recomputed on an RTX 4090 GPU. The blue star denotes AceFF-2-OMol using the TorchMD-NET implementation (with Warp kernels), and the green star denotes the Acellera-TensorNet plugin. Points marked with a $^\dagger$ are MACE models evaluated with randomly initialized weights using architectures corresponding to the released model variants.}
    \label{fig:spice}
\end{figure}

To evaluate out-of-sample generalization within the OMol domain, we benchmarked the MLIPs on the SPICE test set from~\cite{eastman2026}. Although these molecules were strictly held out from all model training, they match the overall chemical scope and complexity of OMOL by focusing on small drug-like organic molecules, peptides, and non-covalent complexes. The benchmark comprises 800 unique systems evaluated across 10 conformations each (8,000 total structures) at the $\omega$B97M-D3BJ/def2-TZVPPD level of theory. It spans four distinct subsets of 200 systems each: small ligands (40–50 atoms), large ligands (70–80 atoms), random-sequence pentapeptides (68–110 atoms), and ligand–amino acid interaction dimers (34–72 atoms). Among these, 617 are neutral and 183 are charged (with formal charges ranging from -4 to +2, predominantly between -1 and +1), offering a direct test of model performance on novel structures within the target biochemical space. The overall MAE values are shown in figure \ref{fig:spice}a plotted against the model parameters. The data is from~\cite{eastman2026}.
We see that AceFF-2 performance is lacking, which is expected because its training set resembles only the small-ligand category. AceFF-2-OMol performs much better, ahead of all other non-OMol25-trained models. Compared to the other OMol25-trained models (UMA, OrbMol, MACE-OMol-0) it performs worse, but its parameter count is one to three orders of magnitude smaller. Note that the other OMol25-trained models are parametrized for all elements in OMol25 rather than just the biomolecular element set we use in AceFF-2-OMol. We also reran the speed benchmarks from~\cite{eastman2026} on an RTX 4090 GPU for consistent comparison. Speed versus accuracy is plotted in Figure~\ref{fig:spice}b. AceFF-2-OMol lies on the Pareto front of speed and accuracy with the TorchMD-NET backend; this advantage is even more pronounced when using the Acellera-TensorNet plugin, which is the fastest model for this system size.

\subsection{Protein--Ligand Interaction Energies (PLA15)}

We evaluated performance on the PLA15 benchmark dataset~\cite{kriz2020pla15}, which comprises 15 protein--ligand active-site complexes alongside their isolated protein and ligand fragments. These complexes contain 283--584 atoms (median 356; 143--270 heavy atoms) and span 23--34\,\AA{} along their longest dimension. Reference interaction energies are computed using a composite scheme: CCSD(T)/CBS for the pairwise-additive component and B3LYP-D3/DZVP for the non-additive component.

Interaction energies are calculated as single-point energies on the deposited geometries via
\begin{equation}
  \Delta E_{\mathrm{int}} = E_{\mathrm{complex}}
    - \left( E_{\mathrm{protein}} + E_{\mathrm{ligand}} \right).
  \label{eq:eint}
\end{equation}

\begin{figure}[htbp]
    \centering
    \includegraphics[width=0.9\linewidth]{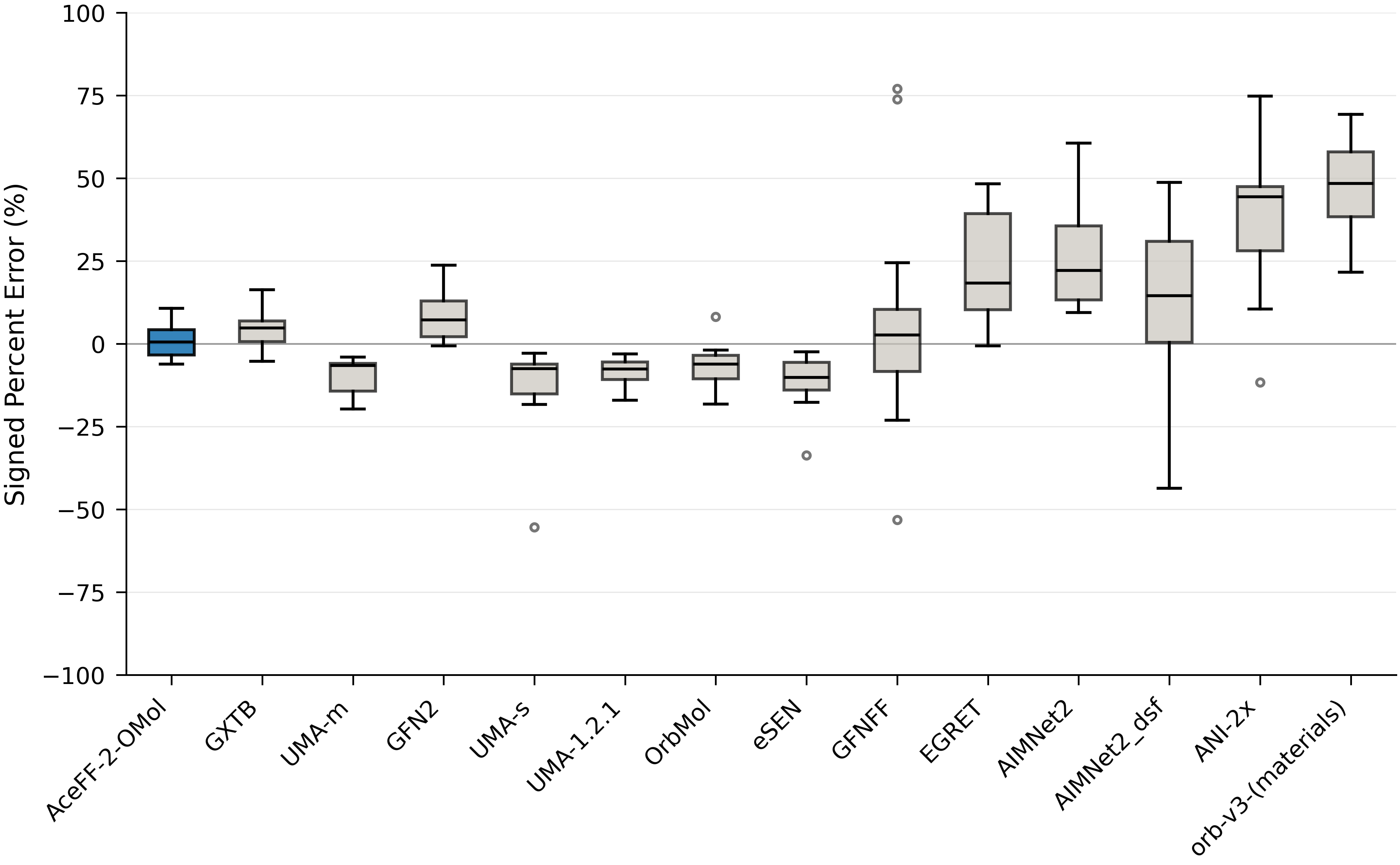}
    \caption{PLA15 benchmark evaluation. Box plots show the distribution of signed percentage errors across all evaluated models.}
    \label{fig:pla15}
\end{figure}

Twelve of the 15 complexes carry a net charge, with 14 featuring a charged protein fragment and 10 a charged ligand. Fragment charges range from $-2$ to $+2$, and the two fragments are oppositely charged in 8 cases. Following the Rowan Scientific benchmarking protocol~\cite{rowanpla15}, we evaluated AceFF-2-OMol and extended the published comparison with two additional models, OrbMol~\cite{orbmol_huggingface} and UMA-1.2.1~\cite{wood2026umafamilyuniversalmodels}; all other values are those reported by Rowan Scientific. We report the distribution of signed percentage errors, defined as $100 \times (\Delta E_{\mathrm{pred}} - \Delta E_{\mathrm{ref}}) / |\Delta E_{\mathrm{ref}}|$, in Figure~\ref{fig:pla15}.

PLA15 lies slightly outside the applicability domain of OMol25, as it contains systems with up to 584 atoms, whereas OMol25 training frames are capped at 350 atoms.

In contrast to results on other benchmarks, the semi-empirical methods g-xTB and GFN2-xTB perform well on PLA15 (Figure~\ref{fig:pla15}). Conversely, most OMol25-trained models systematically overbind, predicting overly strong interaction energies (negative signed errors). Notably, AceFF-2-OMol avoids this overbinding bias, which may be attributed to its inclusion of a long-range learned Coulomb term.

\subsection{Potential energy smoothness}

\begin{figure}[t]
    \centering
    \includegraphics[width=\linewidth]{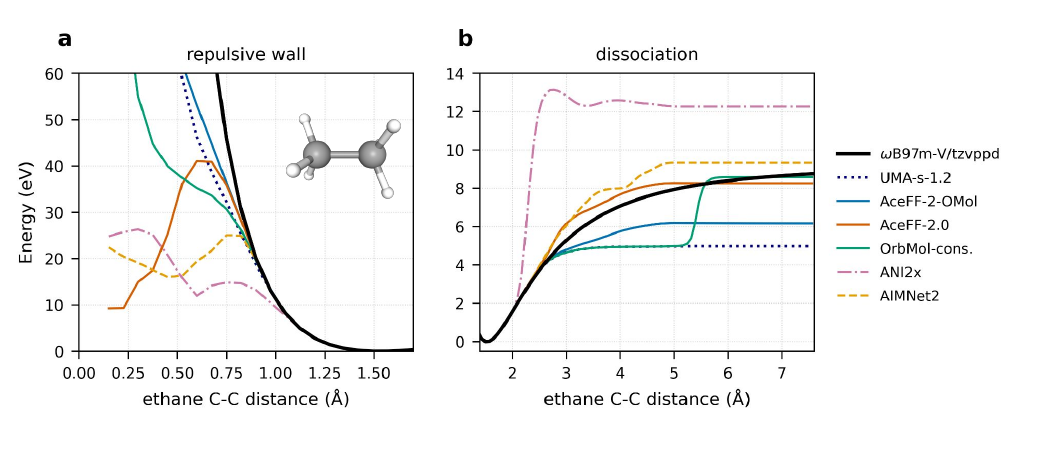}
    \caption{Potential energy of ethane C-C bond length scan for different MLIPs compared to a reference of DFT with $\omega$B97m-V/tzvppd.}
    \label{fig:ethane}
\end{figure}

A smooth potential energy surface is crucial for stable molecular dynamics simulations.
To assess model smoothness, we analyzed the potential energy scan as the C-C bond in an ethane molecule was compressed and extended. We compressed the bond to an unphysically small distance and extended it beyond the bond-breaking regime to examine how the MLIPs behave in these regions outside the training domain. We compared the MLIP results with those from DFT using $\omega$B97m-V/tzvppd. The bond energy scans are plotted in Figure~\ref{fig:ethane}, where they have all been shifted such that the minimum energies are zero.

All models exhibit similar behavior near the minimum. At small distances, we observe the nuclear repulsion wall in the DFT curve; the same feature is reproduced by OrbMol and AceFF-2-OMol, the two models that include an explicit ZBL repulsive term, and by UMA-s-1.2. For the other MLIPs, the nuclear repulsion is modeled entirely by the neural network terms (and the data on which they were trained). The energy barrier in ANI-2x is poorly defined, and AIMNet2 turns over at around 25~eV. AceFF-2 follows the DFT energy up to roughly 30~eV before turning over at about 40~eV, well above any energy sampled in molecular dynamics. At the bond-stretching end of the curve, ANI-2x significantly overestimates the bond strength, UMA-s-1.2 and AceFF-2-OMol underestimate it, and OrbMol exhibits an unphysical step change between 5 and 6~\AA, which is perhaps an artifact of the MLIP cutoffs. AIMNet2 and AceFF-2 have profiles similar to each other and to the DFT reference; these models include a long-range Coulomb term, which enables atom interactions beyond the typical graph cutoff (5~\AA{} to 6~\AA).

\subsection{Inference Speed Scaling with Molecular Dynamics}\label{inference_md}

We compared model evaluation speed by running molecular dynamics on water boxes of increasing size. We used the benchmark systems from \cite{eastman2026} and followed the same simulation protocol. All models were re-benchmarked on identical hardware (an NVIDIA RTX 4090) to ensure a consistent comparison. For the MACE models, we used untrained, randomly initialized weights while preserving the exact architecture and hyperparameters of the published models. In addition to the models evaluated in \cite{eastman2026}, we included ANI-2x as a standard baseline for speed and accuracy. Where supported, models were run using their optimal execution backends: cuAEV for ANI-2x, cuEquivariance for MACE, NVIDIA Warp kernels for TorchMD-NET and AIMNet2, and \texttt{torch.compile}.
 
\begin{figure}[htbp]
    \centering
    \includegraphics[width=0.5\linewidth]{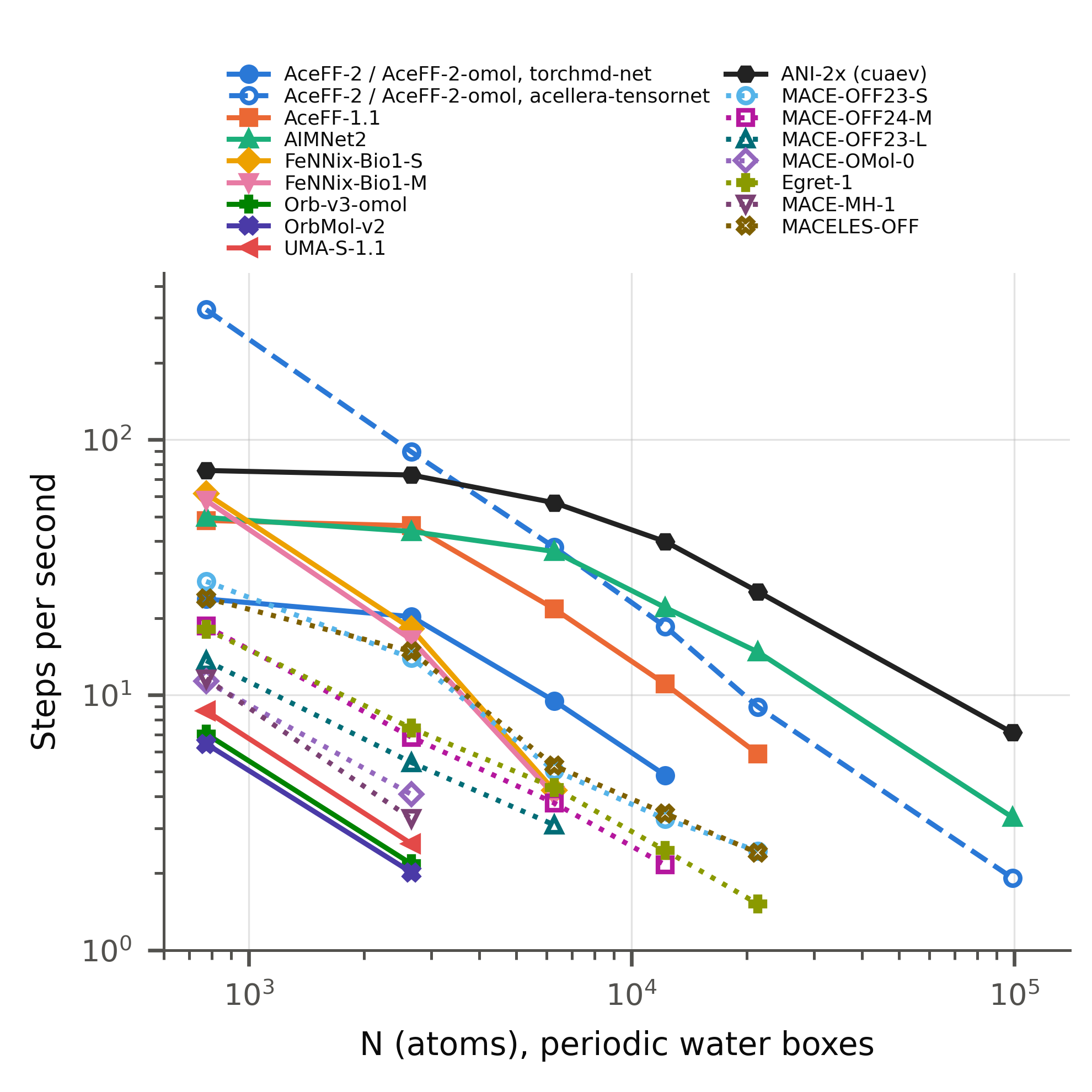}
    \caption{Inference speed in steps per second for different models versus the number of atoms. We use the water box speed benchmarks from~\cite{eastman2026}.}
    \label{fig:inference_speed}
\end{figure}

The throughput (steps per second) for all models as a function of system size is shown in Figure~\ref{fig:inference_speed}. 

There is a clear latency ceiling of approximately 100 steps/second across all models except the Acellera-TensorNet plugin. This bottleneck stems from PyTorch framework overhead and the requirement to compute forces via an automatic differentiation backward pass at each MD step.

Once past the latency-dominated flat region at small $N$, all models scale as expected, exhibiting linear $\mathcal{O}(N)$ scaling due to their local, short-range cutoffs. For the largest systems, ANI-2x is the fastest, followed by AIMNet2 and AceFF-2-omol (Acellera-TensorNet). Note that ANI-2x relies on the cuAEV extension for optimized atomic environment vector calculations, whereas AIMNet2 utilizes NVIDIA Warp kernels for its message-passing convolution layers. A key takeaway is that custom CUDA kernels are essential for achieving competitive MLIP inference speeds.

\begin{figure}[t]
    \centering
\includegraphics[width=\linewidth]{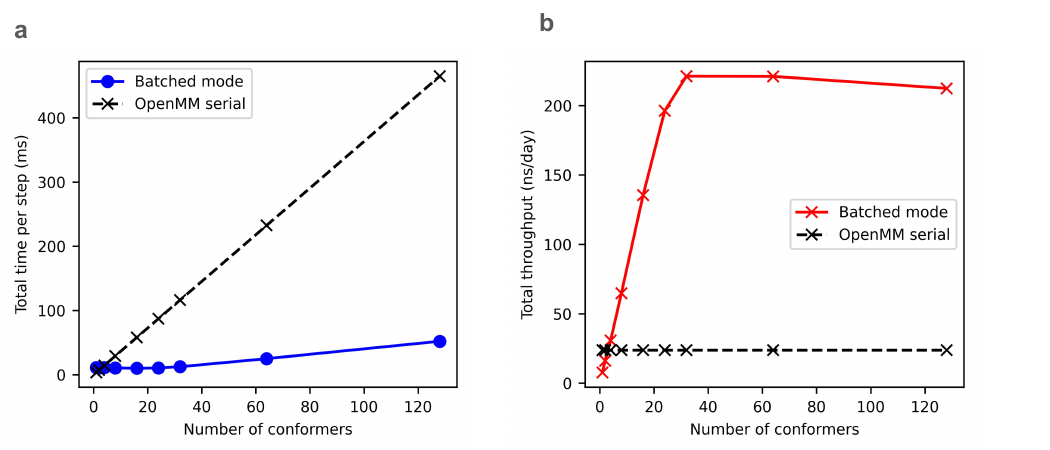}
    \caption{Batched molecular dynamics with AceFF-2. {\bf a.} The total time per step for the corresponding number of conformers in batched mode (blue line) and in serial with OpenMM (black line). {\bf b.} The effective total throughput in ns/day for batched mode (red line) and using OpenMM in serial (black line).}
    \label{fig:batched_md}
\end{figure}

We also tested the speed of batched molecular dynamics. To do this, we selected a small molecule (imatinib, 68 atoms) and generated N conformers. We then ran Langevin dynamics, using the integrator from TorchMD~\cite{doerr2021torchmd}, in a batched manner, such that all N conformers were integrated simultaneously. TorchMD is a molecular dynamics engine entirely written in PyTorch, which allows for easy integration and batching of molecular simulations using MLIPs and the Amber potential. This allowed the AceFF MLIP to be called once per timestep for the entire batch. 
The total time per step is plotted in figure~\ref{fig:batched_md}a for increasing batch size. We compare against the serial OpenMM simulation speed, i.e., the time to sequentially run a timestep for each conformer individually. We also compute the effective total throughput in ns/day for the batched simulations, comparing against the OpenMM simulation throughput of running a single simulation. These results are plotted in figure~\ref{fig:batched_md}b. We see that for 32 conformers there is almost a 10x increase in the total simulation throughput; this is due to the latency limits of running a full forward+backward pass being overcome.

\subsection{MLIP/MM molecular dynamics stability}

\begin{figure}[t]
    \centering
\includegraphics[width=\linewidth]{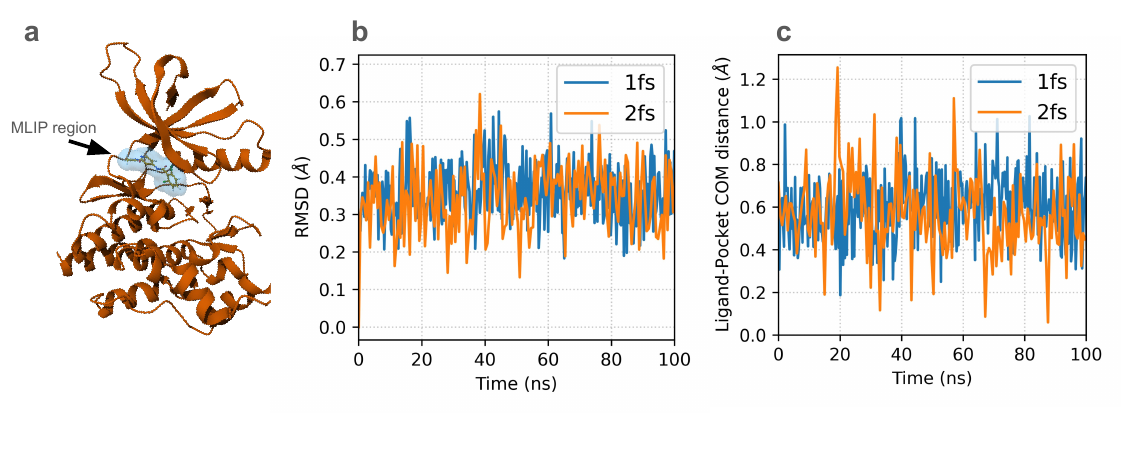}
    \caption{Ligand RMSD of NVT molecular dynamics of protein-ligand complex using ML/MM. {\bf a.} Visualization system; the MLIP region has been highlighted, and water has been omitted for clarity. {\bf b.} The RMSD of the ligand is plotted for the 1fs and 2fs timestep  AceFF-2 simulations in blue and orange, respectively. {\bf c.} The ligand-pocket center of mass distance is plotted for the 1fs and 2fs timestep AceFF-2 simulations in blue and orange, respectively.}
    \label{fig:mm_ml}
\end{figure}

A current use case for MLIPs is in mixed MLIP/MM simulations where a small region is modeled with the MLIP and the rest of the system with a traditional MM forcefield~\cite{Rufa2020.07.29.227959,sabanes2024enhancing, sabanes_zariquiey_quantumbind-rbfe_2025}.
The simplest approach is the mechanical embedding approach, where just the ligand intra-molecular forces are computed with the MLIP and the rest of the system uses the MM force field. This accounts for the corrections due to ligand strain.

To test that AceFF-2 is appropriate for this, we ran a simulation of the ejm31 ligand bound to Tyk2, which is one of the systems studied in~\cite{sabanes_zariquiey_quantumbind-rbfe_2025} and from ~\cite{ross2023maximal}, the complex is shown in figure~\ref{fig:mm_ml}a. We used the OpenMM-ML package~\cite{eastman2023openmm} to run the simulations with hydrogen mass repartitioning (HMR) to 4~amu, temperature of 300K, and a Langevin thermostat. We ran the simulations for 100 ns with timesteps of 1 fs and 2 fs. The ligand RMSDs are plotted in~\ref{fig:mm_ml}b and the center of mass distance between the ligand and protein pocket are plotted in~\ref{fig:mm_ml}c.  We observe that both 2fs and 1fs are stable.

\subsection{Batched conformer minimization}
A common use case of molecular force fields is to energy-minimize molecular conformations. To assess the ability of the MLIPs to minimize a diverse array of molecules relevant for drug discovery, we tested them on the Platinum Diverse dataset~\cite{friedrich2017}, which contains 2859 crystallographic structures of ligand conformations from protein-ligand complexes.

To perform the benchmark, we took the reference conformer from the dataset and minimized it with the MLIP. It is expected that the minimization should converge, and that the reference structure is close to a local minimum of the MLIP force field, such that after minimization, the RMSD between the reference and the minimized structure should be small. We tested AceFF-2, AceFF-2-OMol, AIMNet2, and OrbMol, and excluded ANI-2x due to the large number of charged molecules present in the dataset and elements outside the ANI-2x subset. For this test, we used the ASE LBFGS minimizer and the ASE calculator interfaces of the MLIP models. 

\begin{figure}[t]
    \centering
    \includegraphics[width=\linewidth]{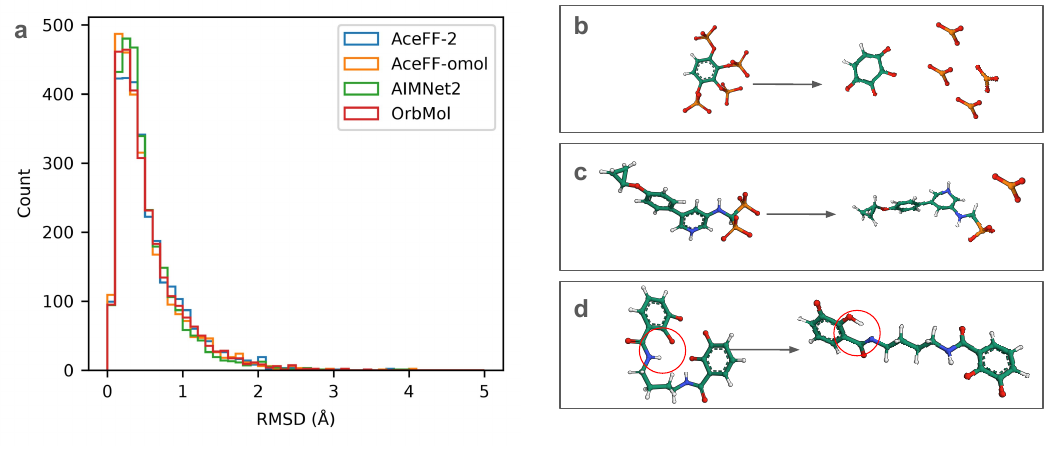}
    \caption{Platinum Diverse geometry optimization benchmark. { \bf a.} The histogram of the RMSDs between the initial reference structure and the geometry-optimized structure for the four tested MLIPs for all conformers in the Platinum Diverse set~\cite{friedrich2017}. { \bf b.} Molecule index 1653  has a net charge of $-8$ due to four phosphate groups. When minimized with AceFF-2 the phosphates are ejected. {\bf c.} Molecule index 1816 has a net charge of $-3$ and two phosphate groups; when minimized with AceFF-2 one of the phosphates is ejected. {\bf d.} Molecule index 1928 has the largest RMSD (3.9~\AA) for AceFF-2; it has a net charge of $-4$, and upon minimization with AceFF-2 there is a change in bond topology as highlighted by the red circles.}
    \label{fig:platinum_minim}
\end{figure}

The number of failed minimizations and the values for the RMSD mean, median, and percentages under 1Å, 2Å, 3Å, and 10Å are shown in table~\ref{tab:platinum_minim}. The histograms of the RMSDs are shown in figure~\ref{fig:platinum_minim}. For consistent comparison, the mean, median, and standard deviation are taken over molecules where all methods succeeded (N=2848).

For AceFF-2, we found that two molecules failed to minimize, resulting in unphysical ``exploded'' structures. These are molecules with index 1653 (name: GVF\_2UVM\_A) and index 1816 (name: JD5\_4NFK\_F). Molecule 1653 has a net charge of -8 due to four -2 phosphate groups; similarly, molecule 1816 has a net charge of -3 and two phosphate groups. Images of these molecules are shown in figure~\ref{fig:platinum_minim}b and ~\ref{fig:platinum_minim}c. The AceFF dataset contains molecules with net charge up to -2 to 2, so these specific molecules are significantly outside the model's domain of knowledge, and it is not surprising that it fails to correctly model them. The next largest RMSD is 3.9 angstroms for molecule 1928; while not exploded, the conformation is in an elongated state compared to the reference, and an H atom has changed bond topology. This molecule has a net charge of -4, so once again outside the training domain. An image of this molecule is shown in figure~\ref{fig:platinum_minim}d. 

For AIMNet2, we found that 8 molecules exploded with RMSDs greater than 10. The molecule indices are [259, 1160, 1468, 1648, 1816, 2189, 2328, 2803], all of which contain the same large net charge and phosphate groups that AceFF struggled with. For AceFF-2-OMol, seven molecules failed with RMSDs greater than 10\,\AA. These are the molecules with index 259
(1MV\_4LPG\_F), 339 (252\_2Z4X\_A), 1160 (B29\_2E98\_A), 1816 (JD5\_4NFK\_F), 2189
(P2H\_4JZB\_A), 2609 (UNR\_3RYE\_A) and 2803 (YS4\_4DEM\_F). All seven are geminal
bisphosphonates, in which two fully deprotonated phosphonate groups sit on the same
carbon atom, and all carry a net charge of $-3$ or $-4$.  OrbMol succeeds at minimizing all conformers.

These conformers are taken from crystal structures of solvated protein--ligand complexes,
in which the phosphonate groups are coordinated to metal ions or to basic active-site
residues, and would be partly protonated at physiological pH. Optimized in the gas phase
as isolated poly-anions, they place four formal negative charges within two bonds of one
another, and the reference geometry is no longer a minimum: the mutual repulsion of the
phosphonate groups drives them apart. AceFF-2-OMol predicts partial charges and adds an
explicit long-range Coulomb term, so it represents this repulsion directly; removing that
term reduces the RMSD of all seven molecules to below 1.7\,\AA. OrbMol uses the total
charge only as a global conditioning input and has no explicit electrostatic term, so it
cannot express the long-range repulsion and instead stays close to the reference geometry.
Its success on these molecules therefore reflects the absence of the interaction rather
than a better description of it.

\begin{table}[t]
  \centering
  \small
  \begin{tabular}{lrrrrrrrr}
    \toprule
    MLIP & mean & std & median & \%$<$1\,\AA & \%$<$2\,\AA & \%$<$3\,\AA & \%$<$10\,\AA & Number failed \\
    \midrule
    AceFF-2 & 0.551 & 0.447 & 0.419 & 86.29 & 98.43 & 99.72 & 99.93 & 2 \\
    AceFF-2-OMol & 0.527 & 0.446 & 0.392 & 87.20 & 98.46 & 99.58 & 99.76 & 7 \\
    AIMNet2 & 0.508 & 0.426 & 0.388 & 89.75 & 98.60 & 99.62 & 99.72 & 8 \\
    OrbMol & 0.534 & 0.433 & 0.400 & 86.92 & 98.64 & 99.86 & 100.00 & 0 \\
    \bottomrule
  \end{tabular}
      \caption{Platinum Diverse geometry optimization benchmark~\cite{friedrich2017}. RMSD (\AA) between the reference crystallographic ligand conformation and the MLIP-minimized structure. Mean, standard deviation, median, and cumulative percentages are computed over the conformers for which all four methods succeeded. ``Number failed'' is the number of conformers with an RMSD above 10~\AA{} (exploded structures).}
    \label{tab:platinum_minim}
\end{table}

\begin{figure}[t]
    \centering
    \includegraphics[width=0.5\linewidth]{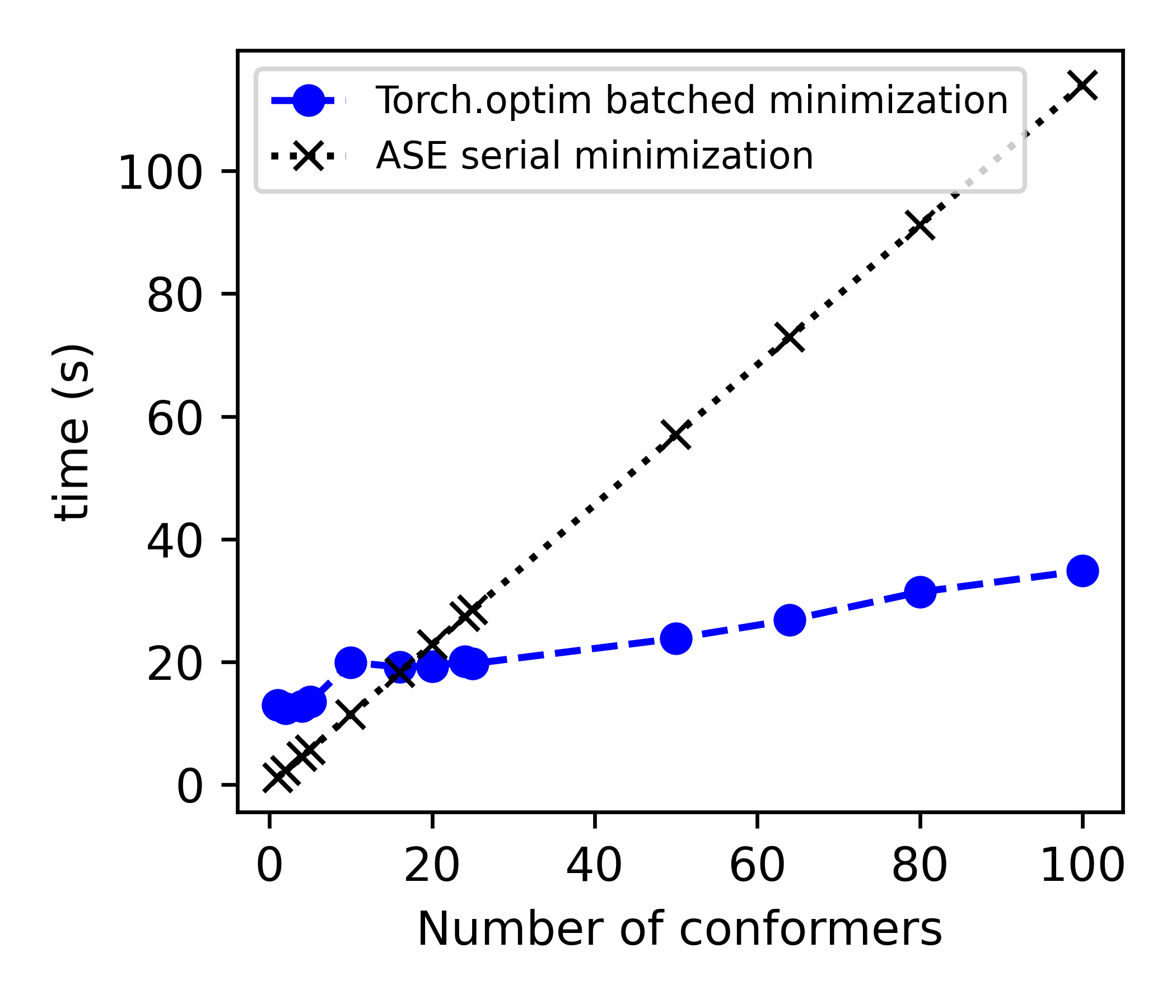}
    \caption{Comparison of batched conformer minimization speed with torch.optim.LBFGS and ASE serial minimization of the same number of conformers of an Imatinib molecule.}
    \label{fig:batched_minim_speed}
\end{figure}

We also assessed the speed of optimizing multiple conformers. The benefit of MLIPs is that batching calculations is trivial using the existing optimizers available in PyTorch. Helpfully, LBFGS, one of the standard methods for geometry optimization, is available. We recorded the time taken to minimize N conformers of Imatinib, a drug molecule with 68 atoms~\cite{pubchem_imatinib_2025}.
We varied N from 1 to 100 and put all conformers into the same batch. For comparison, we timed ASE LBFGS minimization (for the same number of steps, in this case 400) for one conformer, and scaled this linearly by N to get the ASE serial speed baseline. The measured timings are shown in Figure~\ref{fig:batched_minim_speed}. The batched speed is significantly faster for more than 20 conformers. For fewer than 10 conformers, the ASE version is faster because it utilizes torch.compile and CUDA Graphs to minimize execution latency. Conversely, the batched version used with torch.optim does not employ CUDA Graph settings.

\section{Conclusion}
We have presented AceFF-2 and AceFF-2-OMol, two next-generation neural network potentials built on the TensorNet2 architecture and tailored for the efficient and accurate simulation of small drug-like molecules. TensorNet2 extends TensorNet with learned partial-charge features, neutral charge equilibration, and an explicit long-range Coulomb term, which together resolve the poor extrapolation to charged species observed in AceFF-1.x. Both models share this architecture and therefore the same inference cost; they differ in their training data. AceFF-2 was trained on our in-house $\omega$B97M-V/def2-TZVPPD dataset of drug-like molecules containing up to 30 atoms, while AceFF-2-OMol was trained on the subset of OMol25 covering the same medicinal chemistry elements plus Na and K.

We evaluated both models using a comprehensive set of benchmarks, comparing them against other MLIPs, semi-empirical methods, and a classical force field. On torsion scans, strained conformers (Wiggle150), the Schr\"odinger ligand set, and the SPICE test set, AceFF-2 shows a substantial improvement over AceFF-1.0 and is consistently more accurate than AIMNet2, the model closest to it in speed and applicability domain. AceFF-2-OMol improves further on almost every benchmark: it is the most accurate model not trained on the full OMol25 dataset on the SPICE test set, and it approaches the accuracy of the much larger OMol25-trained foundation models (UMA, OrbMol, MACE-OMol-0) with one to three orders of magnitude fewer parameters. On the PLA15 protein--ligand interaction energies, AceFF-2-OMol is the only OMol25-trained model that does not systematically overbind, which could be atrributed to its explicit, learned long-range Coulomb term.

In the current MLIP landscape, ANI-2x~\cite{ANI2x} represents one extreme, offering the fastest speed but limited elemental coverage and a restriction to neutral species. The OMol25-trained foundation models, OrbMol~\cite{orbmol_huggingface,rhodes2025orbv3atomisticsimulationscale,neumann_orb_2024} and UMA~\cite{levine2025openmolecules2025omol25,wood2026umafamilyuniversalmodels}, represent the other extreme, offering high accuracy across most of the periodic table and diverse charge states at a considerably higher computational cost. The AceFF models occupy a position between these extremes on the accuracy--speed Pareto front: they are among the fastest models available for systems below a few thousand atoms, and AceFF-2-OMol in particular approaches the accuracy of the foundation models within the biomolecular element set.

The TensorNet2 architecture further extends the expressivity of TensorNet and remains one of the fastest equivariant MLIPs available. Most of the models we compared against (OrbMol, ANI-2x, AIMNet2) are invariant, a design choice that may have been influenced by the typically higher speed of invariant models~\cite{rhodes2025orbv3atomisticsimulationscale}; the equivariant models we tested, UMA and MACE-MPA-0, are significantly slower. We have not investigated direct force prediction MLIPs (i.e., where forces are output in the forward pass and not by back-propagation) as their inherently non-conservative nature adds significant complexity to their use in stable MD simulations~\cite{bigi2025darkforcesassessingnonconservative}.

Both models support all standard drug-relevant elements, are explicitly trained on charged species, and show strong transferability to larger, out-of-distribution molecules. AceFF-2 produces stable molecular dynamics trajectories for both in-vacuum simulations and hybrid MLIP/MM protein--ligand systems, and AceFF-2-OMol, sharing the same architecture and inference code, can be used as a drop-in replacement in the same workflows.

AceFF-2 and AceFF-2-OMol are available to the community under a permissive license, supported by code and tutorials to enable rapid adoption within both the ASE and OpenMM-ML frameworks. We anticipate that these models will facilitate more accurate and computationally tractable molecular simulations, particularly within drug discovery, by providing a reliable alternative to classical force fields and computationally expensive quantum calculations. Furthermore, the benchmarks established in this work provide the community with a rigorous protocol for testing MLIPs in the regime relevant to small-molecule drug discovery.

\section{Methods}

\subsection{Dataset}
We built an in-house dataset of DFT data for drug-like small molecules, comprising approximately 2 million unique molecules from PubChem~\cite{kim_pubchem_2025} with 12 million conformations. The elements included are  H, B, C, N, O, F, Si, P, S, Cl, Br, and I. Molecule sizes are up to 30 atoms. Charge values include [-2,-1,0,1,2]. The DFT level of theory is $\omega$B97M-V/def2-TZVPPD~\cite{mardirossian__2016}. PySCF~\cite{sun_recent_2020} and the plugin GPU4PySCF~\cite{li2024introducting, wu2024enhancing} was used. The dataset includes energy-minimized structures and high-energy geometries. One reason for creating this dataset was the inadequacy of the existing datasets at the time; notably, they did not contain the higher-energy structures we believed were necessary for stable MLIPs. This dataset was used to train AceFF-2. For AceFF-2-OMOL We used a subset of OMol25~\cite{levine2025openmolecules2025omol25} containing the elements  H, B, C, N, O, F, Si, P, S, Cl, Br, I, Na, and K.

\subsection{Models and force fields}

 We used ANI2x~\cite{ANI2x} from TorchANI~\cite{gao2020torchani}. For all evaluations excluding the OpenMM-based ones, we used the latest TorchANI-2~\cite{pickering_torchani_2025} (version 2.6) For the OpenMM-based simulation, we used an older version (2.2.4,) which is compatible with the NNPOps~\cite{galvelis2023nnp} (version 0.6) package. For OrbMol-conservative~\cite{orbmol_huggingface,rhodes2025orbv3atomisticsimulationscale,neumann_orb_2024} we used version 0.5.5 and all calculations were done via the ASE calculator interface. In the text, we refer to the model as OrbMol-conservative and OrbMol interchangeably. We only ever use the conservative model. For GFN2-XTB\cite{bannwarth_gfn2-xtbaccurate_2019} we used version 22.1 from the xtb-python package and used the ASE calculator interface for all evaluations. For g-XTB\cite{froitzheim_g-xtb_2025} we used the preliminary version 1.1.0 \url{https://github.com/grimme-lab/g-xtb}. We used GAFF\cite{wang2004development,wang2006automatic} version 2.2 via the OpenMM force-fields package and did all evaluations with OpenMM~\cite{eastman2023openmm} version 8.4. For MACE-MPA-0 we used mace-torch version 0.3.14 and the MPA-0 foundation model~\cite{batatia2023foundation}. For UMA-s-1.2 we used the omol task from fairchem-core version 2.17.  We used ASE~\cite{hjorth_larsen_atomic_2017} version 3.26. We used OpenMM-ML version 1.2, OpenMM-Torch version 1.5, and OpenMM version 8.4.

Some models have specific limitations and were therefore excluded from certain tests. ANI2x supports only elements H, C, N, O, S, F, and Cl. g-XTB lacks analytical gradients, so we did not use it for tests requiring force calculations.

\subsection{Torsion scans}
For the Sellers et al torsion scan, we performed relaxed torsion scans as done in \cite{smith_approaching_2019}. We used the GeomeTRIC\cite{wang2016geometry} package to perform geometry optimizations with the constrained torsion angles. We took the CCSD(T)/CBS conformation as the starting point for each optimization. To compare with the coupled cluster reference, we computed the relative energy of the torsion scan compared to the minimum energy along the scan. To report a single number from each torsion scan, we computed the mean absolute error of the scan from the reference CCSD(T)/CBS scan (MAE). 

For the Behara et al torsion scan we followed the protocol and scripts provided in~\cite{behara2024}: We evaluated the single-point energy of each geometry to calculate the torsion scans. For each torsion scan, we then computed the MAE compared to the reference scan as the difference in the relative energies.

\subsection{Molecular dynamics}
For the MLIP/MM molecular dynamics we used a customized version of the OpenMM-ML package along with OpenMM-Torch version 1.5 and OpenMM version 8.4 on an RTX4090. The simulations were run for 100ns with snapshots saved every 10000 steps. The LangevinMiddle integrator was used with a frequency of 1/ps, a temperature of 300K, HMR of 4amu, and a timestep of either 1fs or 2fs.

\subsection{Software and data availability}
TensorNet2 is available under the MIT license at \url{https://github.com/torchmd/torchmd-net}. We have provided a set of example IPython notebooks (which can be run on Google Colab) and a Python package containing code to enable using AceFF as an ASE (Atomic Simulation Environment~\cite{hjorth_larsen_atomic_2017}) calculator and in OpenMM-ML\cite{eastman2023openmm} as an ML potential. The package and example notebooks can be found on the GitHub repository \url{https://github.com/Acellera/aceff_examples}.
The AceFF models can be downloaded from HuggingFace~\cite{aceff2_huggingface,aceff_omol_huggingface}. A tutorial to use the models for MLIP/MM simulations can be found at \url{https://software.acellera.com/acemd/nnpmm.html}.
Acellera-TensorNet is available in the PlayMolecule AI interface \url{https://playmolecule.ai/}.
The dataset for the Wiggle150 benchmark can be found at~\cite{brew_wiggle150_2025}. The dataset for the Sellers et al torsion scan can be found at~\cite{sellers2017comparison}. The structures for the Schrödinger Ligand set can be found at~\cite{ross2023maximal} and \url{https://github.com/schrodinger/public_binding_free_energy_benchmark}. The Behara et al torsion scan data can be found at~\cite{behara2024}. The Platinum Diverse dataset can be found at~\cite{friedrich2017}.

\section*{Author contributions}
S.E.F. and G.D.F. wrote the manuscript. S.E.F. implemented the TensorNet2 architecture. S.E.F., S.D., and A.M. developed the AceFF models and datasets. S.E.F and F.S.Z performed the model benchmarks. A.M. and S.P. provided the warp kernel code.  G.D.F directed the work. 

\section*{Acknowledgments}
The authors thank the volunteers of GPUGRID.net for donating computing time. This project has received funding from the Torres-Quevedo Program from the Spanish National Agency for Research (PTQ2023-012967/AEI/10.13039/501100011033). AM is financially supported by Generalitat de Catalunya’s Agency for Management of University and Research Grants (AGAUR) PhD grant 2025 FI-2-00278.

\section*{Competing interests}
The authors declare the following competing financial interest(s): S.E.F, S.D, F.S.Z, and G.D.F. have a potential conflict of interest due to direct interests in Acellera (See affiliations).

\bibliography{ref}

\end{document}